\documentclass[letterpaper, 10 pt, journal, twoside]{ieeetran}

\IEEEoverridecommandlockouts                              

\usepackage[%
    style=ieee,
    sorting=none,
    natbib=true,  
    backend=biber,
    sortcites=true,
    doi=false,
    url=false,
    mincitenames=1,
    maxcitenames=1, 
    minbibnames=2, 
    maxbibnames=3, 
    hyperref
]{biblatex}

\usepackage[nolist,printonlyused]{acronym}
\usepackage{amsmath, amssymb, amsfonts}

\allowdisplaybreaks
\usepackage{amsthm}
\usepackage{algorithm}
\usepackage{algpseudocode}
\usepackage{balance}
\usepackage{bm}
\usepackage[english]{datetime2}
\usepackage{glossaries}
\usepackage{multicol}
\usepackage{textcomp}

\usepackage{sidecap}
\sidecaptionvpos{figure}{t}

\DeclareMathOperator*{\argmin}{\arg\!\min}

\usepackage[parse-numbers=false]{siunitx}
\usepackage{xcolor}

\makeatletter
\let\MYcaption\@makecaption
\makeatother

\usepackage[font=footnotesize]{subcaption}

\makeatletter
\let\@makecaption\MYcaption
\makeatother

\usepackage{graphicx}
\usepackage{hyperref}
\usepackage{cleveref}
\usepackage{mathtools}

\crefname{figure}{Fig.}{Figs.}
\crefname{equation}{}{}

\crefname{table}{Table}{Tables}
\crefname{algorithm}{Alg.}{Algorithms}
\crefname{section}{Section}{Sections}
\crefname{remark}{Remark}{Remarks}
\def\BibTeX{{\rm B\kern-.05em{\sc i\kern-.025em b}\kern-.08em
    T\kern-.1667em\lower.7ex\hbox{E}\kern-.125emX}}

\usepackage{lineno}

\newcommand\edit[1]{\textcolor{blue}{#1}}

\renewcommand\edit[1]{#1}

\newcommand{\T}{\intercal}

\newcommand{\agent}[1]{\mathcal{A}_{#1}}

\newcommand{\stack}[2]{\mathcal{S}^{#1}_{#2}}
\newcommand{\ctrl}[2]{u^{#1}_{#2}}
\newcommand{\ctrlbm}[2]{\bm{u}^{#1}_{#2}}
\newcommand{\costfn}[1]{J^{#1}}
\newcommand{\hyp}[1]{H_{#1}}

\newcommand{\numplayers}{N}
\newcommand{\numparticles}{N_s}
\newcommand{\numstates}{n}
\newcommand{\numctrls}[1]{m^{(#1)}}
\newcommand{\horizon}{T}

\newcommand{\state}[1]{x_{#1}}
\newcommand{\statebm}[1]{\bm{x}_{#1}}
\newcommand{\cov}[1]{\Sigma_{#1}}
\newcommand{\context}[1]{c_{#1}}
\newcommand{\contextbm}[1]{\bm{c}_{#1}}

\newcommand{\eststate}[1]{z_{#1}}
\newcommand{\eststatebm}[1]{\bm{z}_{#1}}
\newcommand{\estcov}[1]{\Sigma_{#1}}

\newcommand{\procnoise}{W}

\newcommand{\reals}{\mathbb{R}}
\newcommand{\prob}[1]{p\{#1\}}
\newcommand{\probl}[1]{p\left\{#1\right\}}

\newcommand{\GT}{\text{GT}}

\newcommand\extrafootertext[1]{%
    \bgroup
    \renewcommand\thefootnote{\fnsymbol{footnote}}%
    \renewcommand\thempfootnote{\fnsymbol{mpfootnote}}%
    \footnotetext[0]{#1}%
    \egroup
}

\begin{document}

\begin{acronym}
    \acro{RV}{random variable}
    \acro{ILQR}{Iterative Linear Quadratic Regulation}
    \acro{ILQGames}{Iterative Linear-Quadratic Games}
    \acro{SILQGames}{Stackelberg Iterative Linear-Quadratic Games}
    \acro{SLF}{Stackelberg Leadership Filter}
    \acro{LQ}{linear-quadratic}
\end{acronym}

\title{Leadership Inference for Multi-Agent Interactions}

\author{Hamzah I.\ Khan and David Fridovich-Keil
\thanks{Manuscript received: October 26, 2023; Revised:
January 26, 2024; Accepted: February 19, 2024. This paper was recommended for publication by Editor Lucia Pallottino upon evaluation of the Associate Editor and Reviewers' comments. This work was supported by the National Science Foundation under Grant No. 2211548.
The authors ({\tt\small \{hamzah, dfk\}@utexas.edu}) are with the \textit{Department of Aerospace Engineering and Engineering Mechanics}, University of Texas at Austin.}
\thanks{Digital Object Identifier (DOI): see top of this page.}
}

\markboth{IEEE Robotics and Automation Letters. Preprint Version. Accepted February, 2024}{Khan \MakeLowercase{\textit{et al.}}: Leadership Inference for Multi-Agent Interactions}

\maketitle

\begin{abstract}
Effectively predicting intent and behavior requires inferring leadership in multi-agent interactions.
Dynamic games provide an expressive theoretical framework for modeling these interactions. Employing this framework, 
we propose a novel method to infer the leader in a two-agent game by observing the agents' behavior in complex, long-horizon interactions.
We make two contributions. First, we introduce an iterative algorithm that solves dynamic two-agent Stackelberg games \emph{with nonlinear dynamics and nonquadratic costs}, and demonstrate that it consistently converges \edit{in repeated trials}.
Second, we propose the \ac{SLF}, an online method for identifying the leading agent in interactive scenarios based on observations of the game interactions.
We validate the leadership filter's efficacy
on simulated driving scenarios to demonstrate that the \ac{SLF} can draw conclusions about leadership that match right-of-way expectations. \\


\end{abstract}

\newcommand{\LQConvThresholdSILQ}{1.5 \cdot 10^{-2}}
\newcommand{\LQMaxItersSILQ}{50}
\newcommand{\LQMinStepSize}{10^{-2}}
\newcommand{\LQHorizon}{10}
\newcommand{\LQSamplePeriod}{0.02}
\newcommand{\LQTimeSteps}{501}
\newcommand{\LQNumSims}{100}
\newcommand{\LQanglediff}{0.4}
\newcommand{\NonLQNumSims}{100}
\newcommand{\NonLQanglediff}{0.4}
\newcommand{\NonLQConvThresholdSILQ}{1.2 \cdot 10^{-3}}
\newcommand{\NonLQMaxItersSILQ}{3500}
\newcommand{\NonLQMinStepSize}{10^{-2}}
\newcommand{\NonLQHorizon}{10}
\newcommand{\NonLQSamplePeriod}{0.02}
\newcommand{\NonLQTimeSteps}{501}

\vspace{-8pt}
\section{Introduction}
\begin{figure}[!ht]
\centering
\includegraphics[width=0.85\columnwidth]{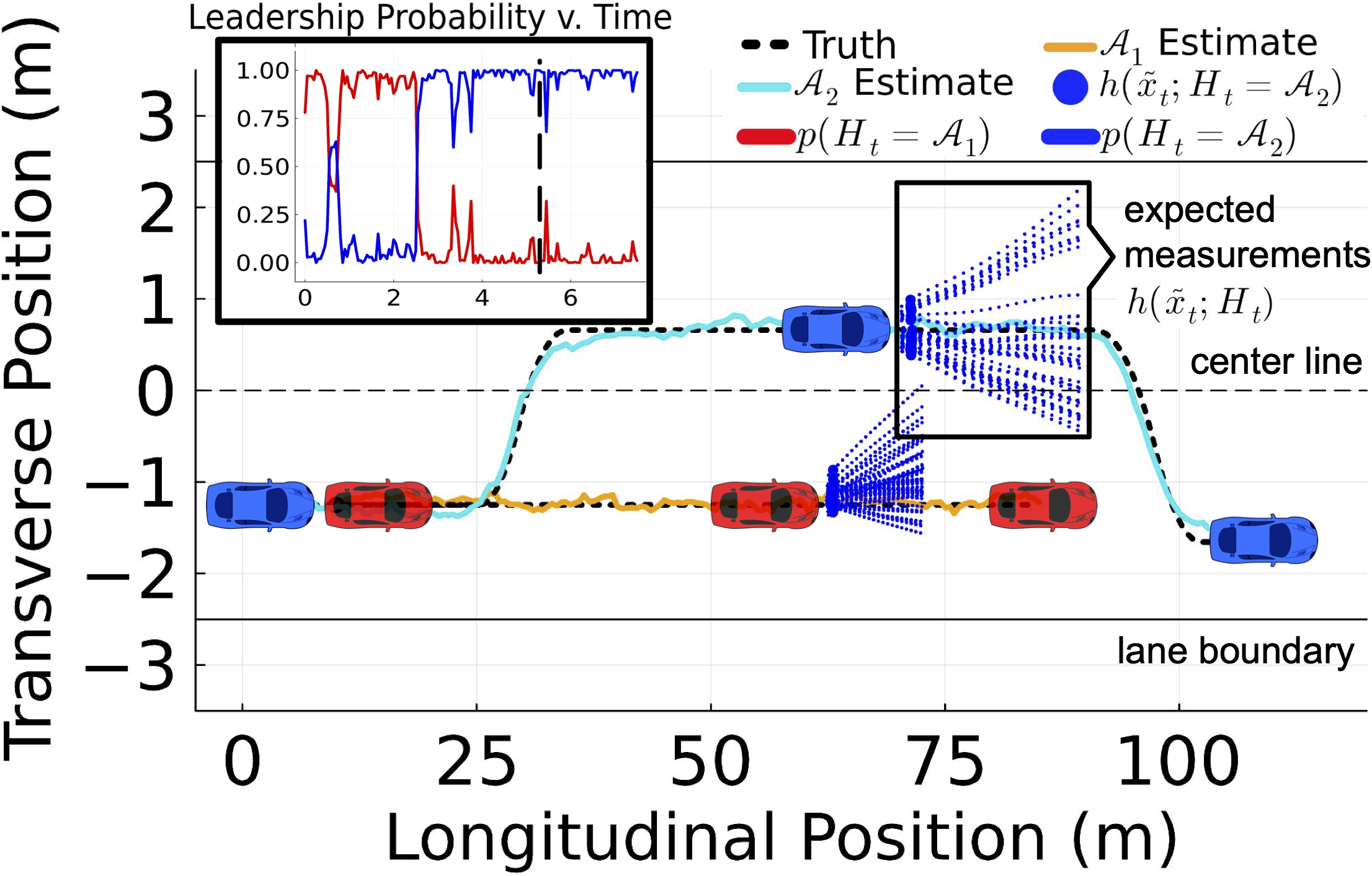}
\caption{Agents $\agent{1}$ (red) and $\agent{2}$ (blue) initially proceed along the same lane of a two-way road at similar speeds. 
While $\agent{2}$ is behind $\agent{1}$, the \ac{SLF} infers that $\agent{1}$ is the leader. During $\agent{2}$'s passing maneuver, the \ac{SLF} captures the leadership probability shifting to $\agent{2}$.
The dashed line in the inset indicates the probabilities at the current time. We display the current expected measurements generated by the measurement model $h$. The blue coloring indicates that most particles in the \ac{SLF} believe $\agent{2}$ is the leader.
\vspace{-12pt}
}
\label{fig:passing-diagram}
\end{figure}

\IEEEPARstart{D}{uring} daily commutes, drivers assert themselves in running negotiations with other road users in order
to reach their destinations quickly and safely. Right-of-way expectations inform these assertions between road users. 
Consider the passing lane shown in \cref{fig:passing-diagram}. Agent $\agent{2}$ (blue) initially follows behind agent $\agent{1}$ (red), and we may intuitively perceive $\agent{1}$ as the leader.
If instead $\agent{2}$ overtakes $\agent{1}$, the scenario seems to imply a reversal of leadership, with $\agent{2}$ in front and $\agent{1}$ behind, as in the inset of \cref{fig:passing-diagram}. However, this intuition is vague and premature. If $\agent{2}$ tailgates $\agent{1}$ or otherwise behaves aggressively, $\agent{1}$ might speed up or yield to $\agent{2}$ out of caution. However, aggressive behavior does not necessarily indicate leadership, as $\agent{1}$ could also react to $\agent{2}$ tailgating by slowing down and relying on the knowledge that $\agent{2}$ will not risk a collision. Here, any simple intuition of the leadership dynamics falls short. Depending on each driver's safety and comfort tolerances, either $\agent{1}$ or $\agent{2}$ may be the leader.
Hence, deciphering leadership dynamics requires understanding common expectations, agent incentives, and other agents' actions.
Successfully doing so can improve autonomous intent and behavior prediction for motion planning, as shown by \cite{tian2022safety}.

We turn to optimal decision making and game theory for tools to analyze interactive scenarios. Stackelberg games \cite{stackelberg1934marktform}, also known as leader-follower games, stand out because they model interactions with clear leadership hierarchies.
In a Stackelberg game,
a leader selects its strategy to influence the follower's response. Each strategy in a Stackelberg solution satisfies leadership conditions that describe how the leader's behavior induces the follower to act.
Additionally, solving Stackelberg games results in trajectories that we can use for model-predictive control.
Using these attractive properties, we propose a leadership inference technique for multi-agent scenarios like that of \cref{fig:passing-diagram}. 
%

%

To this end, we first contribute \ac{SILQGames}, an algorithm for solving \edit{dynamic} Stackelberg games, and we \edit{empirically} show that it converges for games with nonlinear dynamics and general costs.
Second, we propose the \acl{SLF} (\ac{SLF}) to infer leadership over time in interactions based on observations of the agents. We validate that it infers the correct Stackelberg leader in two-agent games and report results 
on simulations of 
driving scenarios.


\section{Related Work}
\label{sec:related-work}


\noindent\textbf{Leadership Inference.} Many prior works develop leadership inference techniques, particularly for robotic swarms and animal sociology.
As an abstract concept, leadership is challenging to measure \cite{garland2018anatomy,strandburg2018inferring}. Leadership models prespecify particular agent(s) as leaders that influence group motion. Swarm applications \cite{deka2021human, singh2021automatic}
often assume the Reynolds flocking model \cite{reynolds1987flocks}. 
Animal sociology applications define leadership models based on principal component analysis \cite{carmi2011mcmc} or stochastic inference \cite{li2020inferring} with hand-selected domain-specific features.
By contrast, we explicitly frame these interactions in terms of optimal decision making and game theory and therefore utilize the Stackelberg leadership model. Defining a Stackelberg game requires prespecifying a leader and solving one produces equilibrium trajectories for each agent. Hence, by associating a particular leader with solution trajectories in a principled manner, Stackelberg leadership allows for modeling leadership
over long-horizon interactions without hand-crafted heuristics.

\noindent\textbf{Stackelberg Games for Motion Planning.}
Recent advances \cite{dextreit2013game,
geary2020resolving} investigate Stackelberg models of leadership for interactive scenarios involving self-driving vehicles.
In particular, \citet{tian2022safety} incorporate leadership as a latent variable by solving open-loop Stackelberg games and comparing expected leader and follower behaviors with observed agent behaviors.
Our method generalizes this underlying approach to Stackelberg leadership by modeling a joint distribution over game state and leadership. We solve feedback Stackelberg games for richer access to leadership information.

\noindent\textbf{Solving Dynamic Games.}
Identifying computationally efficient game-solving techniques with theoretical guarantees of finding equilibria remains an open area of research. 
Most existing game-solving algorithms consider Nash games, which find equilibria for which each actor is unilaterally optimal given fixed opponent strategies.
These algorithms \cite{di2020local,zhu2022sequential,cleac2022algames,bhatt2022efficient,fridovich2020efficientiterative, nakamura2023opinion} generally use Newton-based schemes based on iterative and dynamic programming algorithms that have been widespread for decades \cite{mayne1973differential, bertsekas2012dynamic}. We note two axes on which such approaches differ: first, these approaches solve either open-loop games \cite{di2020local,zhu2022sequential,cleac2022algames,bhatt2022efficient} or feedback games \cite{di2020local,fridovich2020efficientiterative,nakamura2023opinion}. Second, these algorithms either reduce the game to a simpler problem \cite{bhatt2022efficient} or directly solve the game \cite{di2020local,zhu2022sequential,cleac2022algames,fridovich2020efficientiterative,nakamura2023opinion}. In particular, \citet{fridovich2020efficientiterative} introduce \ac{ILQGames}, an iterative method that approximates solutions to nonlinear dynamic, nonquadratic cost feedback Nash games by repeatedly solving \ac{LQ} approximations until convergence. Convergence analysis of these methods is subtle, as described in depth by \citet{laine2021computation}.
Our work is closely related to \cite{nakamura2023opinion}, which uses ILQ schemes to solve for feedback Stackelberg equilibria.
Both our work and \cite{nakamura2023opinion} utilize similar approaches as \cite{fridovich2020efficientiterative} to solve \edit{for feedback Stackelberg equilibria, a different solution concept than (1) single-agent optima found by \ac{ILQR} or DDP and (2) feedback Nash equilibria found by \cite{fridovich2020efficientiterative}.}


\section{Problem Formulation}
\label{sec:problem-formulation}

\edit{Let $\numplayers=2$} agents, $\agent{1}$ and $ \agent{2}$ (e.g., vehicles), operate in a shared \edit{$\numstates$-dimensional} state space with state $\state{t} 
$ at each time $t \in \mathbb{T} \equiv \{1, 2, \ldots, \horizon\}$ and sampling period $\Delta t$. Agent $\agent{i}$ has controls $\ctrl{(i)}{t} \in \mathbb{R}^{\numctrls{i}}$\!.
The state evolves according to
\vspace{-6pt}
\begin{equation}
\label{eq:state-evolution-ode}
\state{t+1} = f_t\left(\state{t}, \ctrl{(1)}{t}, \ctrl{(2)}{t}\right).
\vspace{-6pt}
\end{equation}
\edit{We denote sequence of states $\statebm{1:\horizon} = (\state{1}, \state{2}, \ldots, \state{\horizon})$ and $\ctrlbm{(i)}{1:\horizon} = (\ctrl{(i)}{1}, \ctrl{(i)}{2}, \ldots, \ctrl{(i)}{\horizon})$ as the sequence of $\agent{i}$'s controls.
}
We assume that $f_t$ is continuous and continuously differentiable in $\state{t}, \ctrl{(1)}{t}, \ctrl{(2)}{t}$.
$\agent{i}$'s objective,
\vspace{-6pt}
\begin{equation}
\label{eq:player-objectives}
\costfn{(i)}\left(\statebm{1:\horizon}, \ctrlbm{(1)}{1:\horizon}, \ctrlbm{(2)}{1:\horizon}\right)
\equiv \sum_{t=1}^\horizon g^{(i)}_t\left(\state{t}, \ctrl{(1)}{t}, \ctrl{(2)}{t}\right),
\vspace{-6pt}
\end{equation}
describes its preferences in a given scenario. We model the objective \cref{eq:player-objectives} as the sum of stage costs $g^{(i)}_t$, assumed to be twice differentiable in $\state{t}, \ctrl{(1)}{t}, \ctrl{(2)}{t}$.
Each agent $\agent{i}$ minimizes its objective with respect to its controls $\ctrlbm{(i)}{1:\horizon}$.


\subsection{Background: Feedback Stackelberg Games}
\label{ssec:background-stackelberg-games}

\newcommand\leader{\edit{L}}
\newcommand\follower{\edit{F}}

Stackelberg games model leadership as a mismatch of information.
Intuitively, the leader $\agent{\leader}$ commits to a strategy and communicates it to the follower $\agent{\follower}$. Given this relationship, the leader carefully selects its strategy in order to influence the follower. 

Formally, a Stackelberg equilibrium $\! \{ \ctrlbm{(\leader*)}{1:\horizon}\!, \ctrlbm{(\follower*)}{1:\horizon}(\ctrlbm{(\leader*)}{1:\horizon}) \}$ is a tuple of optimal control trajectories for both agents.
The function $\ctrlbm{(\follower*)}{1:\horizon}(\ctrlbm{(\leader)}{1:\horizon})$ highlights that $\agent{\follower}$'s optimal strategy depends on the leader's (possibly non-optimal) chosen strategy.
%
Using an abuse of notation, we omit the state argument of \edit{$\agent{i}$'s} objective $\costfn{(i)}$, 
and define $\bm{\gamma}(\ctrl{(i)}{t}) \equiv [\ctrlbm{(i)}{1:t-1}, \ctrl{(i)}{t}, \ctrlbm{(i*)}{t+1:\horizon}]$, 
\edit{containing arbitrary controls from time $1$ to $t-1$, control $\ctrl{(i)}{t}$ passed as a parameter, and an equilibrium strategy $\ctrlbm{(i*)}{t+1:T}$ from time $t+1$ to $T$. We note that the game dynamics ensure that $\ctrlbm{(i*)}{t+1:T}$ is implicitly a function of the state.}
We define the set of all optimal follower responses at time $t$, $U^{(\follower*)}_{t}\left(\ctrl{(\leader)}{t}\right) \subset \mathbb{R}^{\numctrls{\follower}}$, as
\vspace{-6pt}
\begin{equation}
\label{eq:optimal-follower-response-set}
U^{(\follower*)}_{t}\left(\ctrl{(\leader)}{t}\right) \equiv \argmin_{\ctrl{(\follower)}{t}} \costfn{(\follower)}\left( \bm{\gamma}\left( \ctrl{(\leader)}{t} \right)\!, \bm{\gamma}\left(\ctrl{(\follower)}{t}\right) \right).
\vspace{-6pt}
\end{equation}
We assume $|U^{(\follower*)}_{t}(\ctrl{(\leader*)}{t})| = 1$, i.e., that an \emph{optimal} leader strategy results in a unique optimal follower response at each time $t$.
Under this assumption, the set of control trajectories for all agents forms a \emph{feedback Stackelberg equilibrium}
if, at every time $t \in \mathbb{T}$, the optimal trajectories satisfy
\begin{flalign}
\costfn{(\leader)}\left(\bm{\gamma}\left(\ctrl{(\leader*)}{t}\right), \bm{\gamma}\left(\ctrl{(\follower*)}{t}\right) \right) &= \label{eq:stack-leader-equilibria-condition}  \\
\min_{\ctrl{(\leader)}{t}}\max_{\ctrl{(\follower)}{t} \in U^{(\follower*)}_{t}\left(\ctrl{(\leader)}{t}\right)}
&\costfn{(\leader)}\left( \bm{\gamma}\left(\ctrl{(\leader)}{t}\right), \bm{\gamma}\left(\ctrl{(\follower)}{t}\right)  \right). \notag
\end{flalign}
$$\vspace{-16pt}$$
\edit{
Since the follower knows the leader's controls at time $t$, \cref{eq:optimal-follower-response-set} 
ensures that the follower produces a best response at time $t$. 
Next, \cref{eq:stack-leader-equilibria-condition} ensures that the leader's strategy 
guides the follower towards its least bad option for the leader at time $t$.}
%

Stackelberg games are generally \emph{non-cooperative}, meaning that agents do not coordinate but plan based on observations of the game state. Agents in open-loop games observe only the initial game state, whereas
in \emph{feedback} games, agents adjust their control inputs after observing the state at each time step, producing complex, temporally-nested game constraints \cref{eq:optimal-follower-response-set,eq:stack-leader-equilibria-condition}.
\Ac{LQ} Stackelberg games have analytic solutions given strictly convex costs \cite[Eq. 7.14-15]{bacsar1998dynamic}.

We denote $\stack{i}{\horizon}(\state{t})$ as the $\horizon$-horizon Stackelberg game solved from state $\state{t}$ with leader $\agent{i}$. For a more detailed treatment of Stackelberg equilibria and solving \ac{LQ} Stackelberg games, refer to Ba\c{s}ar and Olsder \cite[Ch. 3, 7]{bacsar1998dynamic}.

\subsection{Stackelberg Leadership Filtering}
\label{ssec:slf-formulation}

We seek to describe a filter that identifies a leadership belief for $\agent{i}$ based on observations. To this end, we define $\hyp{t} \in \{1, 2\}$ to be a binary \ac{RV} indicating the leader at time $t$. Next, we state our assumptions about the game's observability. 
We assume state $\state{t}$ is observable via noisy measurement
$\eststate{t} \sim \mathcal{N}(h(\state{t}; \hyp{t}), \estcov{t})$ with 
known covariance matrix $\estcov{t} \succ 0$ and measurement model $h$. We also assume that 
control inputs $\ctrl{(i)}{t}$ for each agent $\agent{i}$ are directly observable.
%
Next, recall that each agent has an objective that describes its preferences. For this work, we assume all agent objectives $\{ \costfn{(i)} \}$ are known a priori.
In general settings, we note that techniques exist \cite{peters2021inferring,mehr2023maximum,liu2022learning} to infer agent objectives from noisy observations, though further work may be required to confirm the computational tractability of simultaneously inferring leadership and objectives. 
%
We 
define the leadership belief for $\hyp{t}$ as $b(\hyp{t})\! =\! \prob{\hyp{t} | \eststatebm{1:t}}$.



\newcommand\belief[1]{b(\hyp{#1})}
\newcommand\beliefx[1]{b(\state{#1})}
\newcommand\beliefboth[1]{b(\state{#1}, \hyp{#1})}
\newcommand\beliefc[1]{b(\context{#1})}
\newcommand\particle[1]{\tilde{x}_{#1}}
\newcommand\particlek[2]{\particle{#1}^{#2}}
\newcommand\hypk[2]{\hyp{#1}^{#2}}
\newcommand\contextk[2]{\context{#1}^{#2}}
\newcommand\contextkbm[2]{\contextbm{#1}^{#2}}
\newcommand\statemeas[1]{\eststate{#1}}
\newcommand\statemeasbm[1]{\eststatebm{#1}}

\section{Inferring Leadership}
\label{sec:filtering-leadership}
\begin{figure*}
\centering
\begin{minipage}{0.61\linewidth}
    \vspace{6pt}
    \includegraphics[width=\columnwidth]{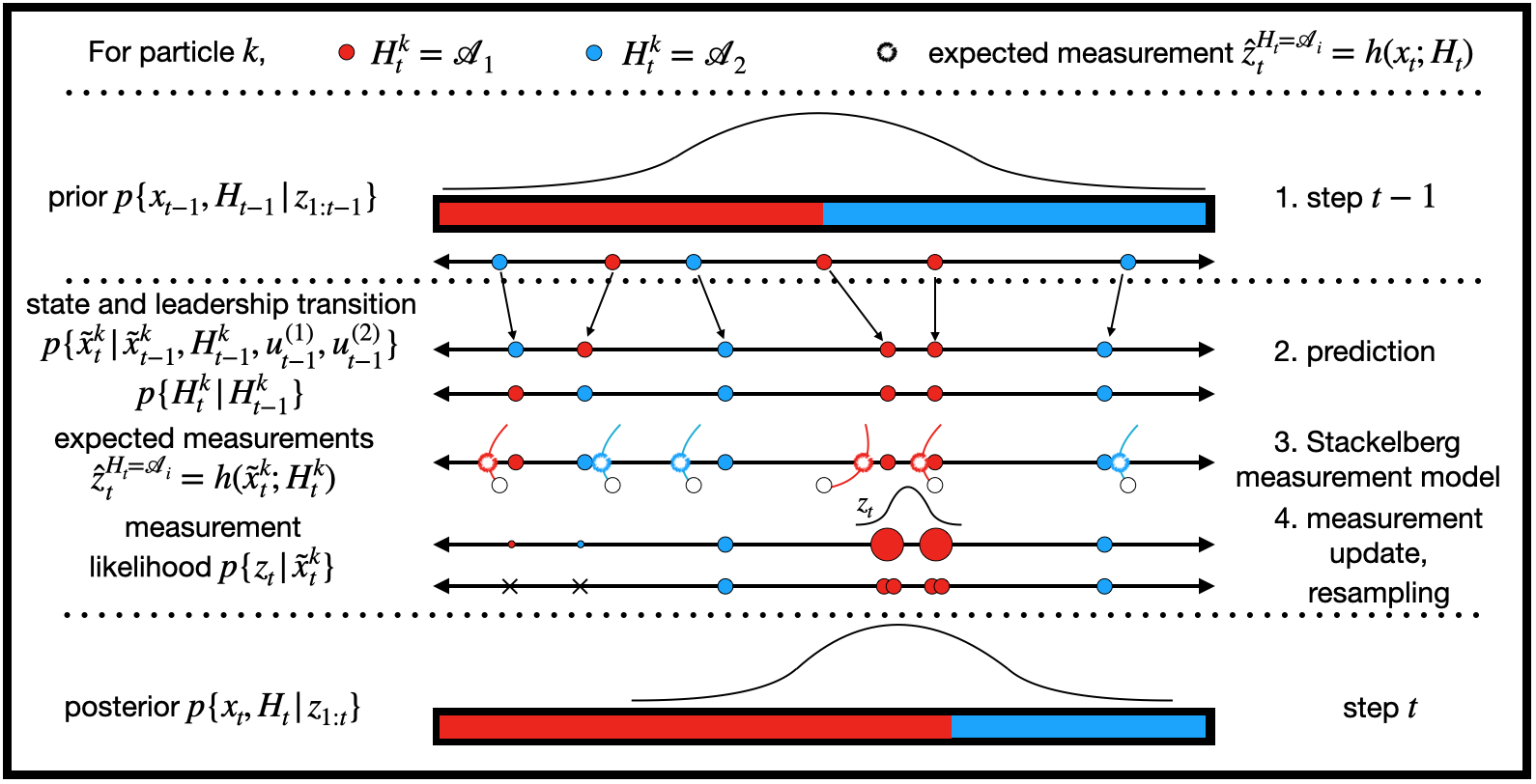}
\end{minipage}
\hfill
\begin{minipage}{0.37\linewidth}
\caption{\captionsize Each particle in the Stackelberg leadership filter has context $\contextk{t}{k} = [\particlek{t}{k}, \hypk{t}{k}]^\T$, where continuous RV $\state{t} \in \mathbb{R}^n$ describes the state and discrete RV $\hyp{t} \in \{1, 2\}$ indicates the leader.
1. At $t-1$, we have a prior distribution over the filter context. For $\hyp{t-1}$, the prior is Bernoulli distributed.
2. The continuous state transitions according to game dynamics $f_{t-1}$.
Leadership state evolves stochastically based on a two-state Markov chain.
3. We play a Stackelberg game from each particle's previous state and extract the game state at the current time $t$ as the expected measurement.
4. The algorithm uses a standard particle filter measurement update \citep[Ch. 4]{thrun2002probabilistic}. Resampling eliminates unlikely particles and reweights the particle set towards those that are similar to the measurement.
Finally, we marginalize over the continuous state and produce a probability of leadership.
%
}
\label{fig:method-block-diagram}
\end{minipage}
\vspace{-15pt}
\end{figure*}

We propose \acl{SILQGames} (\ac{SILQGames}), which iteratively solves nonlinear dynamic, general cost (non-\ac{LQ}) Stackelberg games with continuous and differentiable dynamics and costs.
We use \ac{SILQGames} in the \acl{SLF} (\ac{SLF}, \cref{fig:method-block-diagram}) as part of the Stackelberg leadership model. Our method infers the leading agent of a two-agent interaction from observations.

\newcommand{\dx}[2]{\delta \state{#1}^{#2}}
\newcommand{\dxbm}[2]{\delta \statebm{#1}^{#2}}
\newcommand{\du}[3]{\delta \ctrl{(#1), #3}{#2}}
\newcommand{\dubm}[3]{\delta \ctrlbm{(#1), #3}{#2}}

\newcommand\variterations{\edit{s}}

\subsection{Iteratively Solving Stackelberg Games}
\label{ssec:solving-stackelberg-game}
At a high level, \ac{SILQGames} (\cref{alg:silq-games})
iteratively solves \ac{LQ} approximations of Stackelberg games (\cref{alg:lines:approx1,alg:lines:approx2,alg:lines:approx3,alg:lines:approx4,alg:lines:approx5}), updates the control trajectories using the solutions to these approximated games (\cref{alg:lines:update1}), and terminates if the updated trajectory satisfies a convergence condition (\cref{alg:lines:conv1,alg:lines:conv2,alg:lines:conv3}). Upon successful convergence, the resulting trajectory constitutes an approximate Stackelberg equilibrium.
This type of approach also yields approximate equilibrium solutions in the Nash case, although establishing precise error bounds remains an open problem \citep{laine2021computation}. We expect a similar result for \ac{SILQGames}, though it is beyond the scope of this work.
%
%

\begin{algorithm}
\caption{\acl{SILQGames}} \label{alg:silq-games}
\textbf{Input:} leader $\agent{\leader}$, initial state $\state{1}$, nominal strategies $\left\{ \! \ctrlbm{(i), 0}{1:\horizon} \! \right\}$ \\
\textbf{Output:} converged strategies $\left\{ \ctrlbm{(i), \variterations-1}{1:\horizon} \right\}$
\begin{algorithmic}[1]
\State $\statebm{1:\horizon}^0 \gets \text{applyGameDynamics}\left(\state{1}, \left\{ \ctrlbm{(i), 0}{1:\horizon} \right\} \right)$ \label{alg:lines:init-state-rollout}
\State $\alpha_{\variterations} \gets \alpha_1$ \label{alg:lines:init-step-size}
\For{\emph{iteration} $\variterations = 1, 2, \ldots, M_{\text{iter}}$} \label{alg:lines:new-iter-loop}
    \State $\bm{F}_{1:\horizon} \equiv \{ A, B^{(i)} \}_{t=1:\horizon}$ \hspace{2.725cm} (\ref{eq:linearize-dyn-1}, \ref{eq:linearize-dyn-2}) \label{alg:lines:approx1}\\$\hspace{1.3cm} \gets \text{linearizeDynamics} \left(\statebm{1:\horizon}^{\variterations-1}, \left\{ \ctrlbm{(i), \variterations-1}{1:\horizon} \right\}\right)$ \label{alg:lines:approx2}
    \State $\bm{G}_{1:\horizon} \equiv \{ Q^{(i)}, q^{(i)}, R^{ij}, r^{ij} \}_{t=1:\horizon}$ \hspace{0.2cm} (\ref{eq:quadraticize-costs-1}, \ref{eq:quadraticize-costs-2}, \ref{eq:quadraticize-costs-3}, \ref{eq:quadraticize-costs-4}) \label{alg:lines:approx3}\\ \hspace{1.225cm} $\gets \text{quadraticizeCosts}\left(\statebm{1:\horizon}^{\variterations-1}, \left\{ \ctrlbm{(i), \variterations-1}{1:\horizon} \right\}\right)$ \label{alg:lines:approx4}
    \State $\bm{P}^{(i),\variterations}_{1:\horizon}, \bm{p}^{(i),\variterations}_{1:\horizon} \gets \text{solveLQStackelberg}(\{ \bm{F}_{1:\horizon}, \bm{G}_{1:\horizon} \})$ \label{alg:lines:approx5}
    \State $\statebm{1:\horizon}^{\variterations}, \left\{ \ctrlbm{(i), \variterations}{1:\horizon} \right\} \gets \text{stepToward}( \bm{P}^{(i), \variterations}_{1:\horizon}, \bm{p}^{(i), \variterations}_{1:\horizon}, \alpha_{\variterations})$ \hspace{-0.075cm} \cref{eq:silq-update-eq} \label{alg:lines:update1}
    \If{$\|\statebm{1:\horizon}^{\variterations} - \statebm{1:\horizon}^{\variterations-1}\|_\infty \leq \tau$} \label{alg:lines:conv1}
        \State \textbf{return} $\statebm{1:\horizon}^{\variterations-1}, \left\{ \ctrlbm{(i), \variterations-1}{1:\horizon} \right\}$ \label{alg:lines:conv2}
    \EndIf \label{alg:lines:conv3}
    \State $\alpha_{\variterations+1} \gets \max(\alpha_{\min}, \beta \alpha_{\variterations})$ \label{alg:lines:step-size-1}
\EndFor
\end{algorithmic}
\end{algorithm}
\noindent\textbf{Inputs.} 
\ac{SILQGames} accepts an initial state $\state{1}$ and a leader $\agent{\leader}$.
It accepts a set of all agents' nominal control trajectories $\{ \ctrlbm{(i), \variterations=0}{1:\horizon} \}$.
We produce a nominal state trajectory $\statebm{1:\horizon}^0$ by applying the nominal controls from $\state{1}$ (\cref{alg:lines:init-state-rollout}).

\noindent\textbf{\ac{LQ} Game Approximation.} At each iteration $\variterations$, 
we first linearize the dynamics (\cref{alg:lines:approx1,alg:lines:approx2}) and take second-order Taylor series approximations of the costs (\cref{alg:lines:approx3,alg:lines:approx4}) about the previous iteration's state and control trajectories, $\statebm{1:\horizon}^{\variterations-1}, \ctrlbm{(1),\variterations-1}{1:\horizon}, \ctrlbm{(2),\variterations-1}{1:\horizon}$:

\begin{minipage}{0.95\columnwidth}
\begin{subequations}
\label{eq:lq-approximation-for-silq}
\begingroup
\setlength{\columnsep}{3pt} 
\begin{multicols}{2}
\vspace*{-17pt}
\begin{flalign}
    \hspace{-15pt}A_t &= \nabla_{\state{}} f_t, \label{eq:linearize-dyn-1} \\
    \hspace{-15pt}Q^{(i)}_t &= \nabla^2_{\state{}\state{}}{g^{(i)}_t}, \label{eq:quadraticize-costs-1} \\
    \hspace{-15pt}R^{ij}_t &= \nabla^2_{\ctrl{(j)}{}\ctrl{(j)}{}}{g^{(i)}_t}, \label{eq:quadraticize-costs-2}
\end{flalign}
\columnbreak
\begin{flalign}
    \hspace{-5pt}B^{(i)}_t &= \nabla_{\ctrl{(i)}{}} f_t, \label{eq:linearize-dyn-2} \\
    \hspace{-5pt}q^{(i)}_t &= \nabla_{\state{}}{g^{(i)}_t}, \label{eq:quadraticize-costs-3} \\
    \hspace{-5pt}r^{ij}_t &= \nabla_{\ctrl{(j)}{}}{g^{(i)}_t}. \label{eq:quadraticize-costs-4}
\end{flalign}
\end{multicols}
\endgroup
\end{subequations}
\end{minipage}
$$\vspace{-27pt}$$
%
%
\noindent We define the state and control variables for our \ac{LQ} game approximation as deviations from the previous state and control trajectories: 
$\dxbm{1:\horizon}{\variterations} = \statebm{1:\horizon}^{\variterations} - \statebm{1:\horizon}^{\variterations-1}$ and $\delta\bm{u}^{(i), \variterations}_{1:\horizon} = \bm{u}^{(i), \variterations}_{1:\horizon} - \bm{u}^{(i), \variterations-1}_{1:\horizon}$.
We then approximate the game as an \ac{LQ} problem with linear dynamics and quadratic costs
\vspace{-4pt}
\begin{subequations}
\label{eq:taylor-approximations}
\begin{flalign}
&\hspace{16pt}\dx{t+1}{\variterations} \approx A_t \dx{t}{\variterations} + \sum_{i\in \{1, 2\}} B^{(i)}_t \du{i}{t}{\variterations} \label{eq:approx-dynamics-linearization}, \\
&g^{(i)}_t\left(\cdot, \cdot, \cdot\right) \approx 
g^{(i)}_t\left(\state{t}^{\variterations-1}, \ctrl{(1), \variterations-1}{t}, \ctrl{(2), \variterations-1}{t}
\right) + \frac{1}{2} \|\dx{t}{\variterations} \|^2_{Q^{(i)}_t}
\nonumber \\
    &\hspace{0.4cm}+
    q^{(i)\T}_t \dx{t}{\variterations}
    + \sum_{j=1}^{N} \left(
    \frac{1}{2} \|\du{j}{t}{\variterations} \|^2_{R^{ij}_t} + r^{ij\T}_t
    \du{j}{t}{\variterations} \right), \label{eq:approx-cost-quadraticization}
\end{flalign}
\end{subequations}
$$\vspace{-18pt}$$
\noindent where $\|\cdot\|_{M}$ is an induced matrix norm. We exclude mixed partials $\nabla_{\state{}\ctrl{(i)}{}}, \nabla_{\ctrl{(i)}{}\ctrl{(j)}{}}$ due to their rarity in cost structures of relevant applications, but they can be included if needed.
 
In practice, $Q^{(i)}_t$ and $R^{ij}_t$ 
may not be positive definite. Recall that \ac{LQ} Stackelberg games have unique global solutions given strictly convex costs. Thus, we enforce positive definiteness, and thus convexity, in the quadratic cost estimates by adding a scaled identity matrix $\nu I$ to all $Q^{(i)}_t$ and $R^{ij}_t$ terms. This addition increases each eigenvalue by $\nu \in \mathbb{R}_+$ \cite[Ch. 3]{nocedal1999numerical}, so a sufficiently large choice of $\nu$ guarantees convexity. Finally, we solve the \ac{LQ} game analytically (\cref{alg:lines:approx5}) \cite[Eq. 7.14-15]{bacsar1998dynamic}.

\noindent\textbf{Strategy Update.} After approximating the game as \ac{LQ} and solving it,
we update the control strategy (\cref{alg:lines:update1}). The analytic solution to the \ac{LQ} game consists of gain and feedforward terms $\bm{P}^{(i), \variterations}_{1:\horizon}, \bm{p}^{(i), \variterations}_{1:\horizon}$ constituting an affine feedback control law that produces strategy $\delta\hat{u}^{(i),\variterations}_{t}
= -P^{(i),\variterations{}}_t \delta\state{t}^{\variterations{}} - p^{(i),\variterations{}}_t$.
Following standard procedures in \ac{ILQR} \cite{mayne1973differential}, we define update rule
\vspace{-4pt}
\begin{equation}
\label{eq:silq-update-eq}
\ctrl{(i), \variterations}{t} = \ctrl{(i), \variterations-1}{t} - P^{(i),\variterations}_{t} \dx{t}{\variterations} - \alpha_{\variterations} p^{(i),\variterations}_{t},
\end{equation}
$$\vspace{-24pt}$$
where $\alpha_{\variterations} \in (0, 1]$ is an iteration-varying step size parameter. 
As $\alpha_{\variterations}$ approaches 0, the new iterate $\ctrl{(i), \variterations}{t}$ approaches the previous iterate $\ctrl{(i), \variterations-1}{t}$. Likewise, as $\alpha_{\variterations}$ approaches 1, we adjust our previous iterate by the full step $\delta\hat{u}^{(i),\variterations}_{t}$.
In single-agent settings, methods like \ac{ILQR} commonly apply a line search for step size selection. However, this approach requires a detailed description of complex, temporally-nested feedback game constraints \cref{eq:optimal-follower-response-set,eq:stack-leader-equilibria-condition}.
Instead, \ac{SILQGames} decays the step size (\cref{alg:lines:step-size-1}) with configurable decay factor $\beta \in (0, 1)$ and minimum step size $\alpha_{\min}$.
Initial step size $\alpha_1 = 1$ unless otherwise specified (\cref{alg:lines:init-step-size}).


\noindent\textbf{Convergence Criterion.} 
Optimization algorithms commonly use first-order optimality conditions \cite[Ch. 12]{nocedal1999numerical} to test for convergence, and incorporating a line search guarantees monotone improvement in such a convergence metric.
As with a line search, however, using first-order optimality conditions becomes unwieldy due to the feedback game constraints \cref{eq:optimal-follower-response-set,eq:stack-leader-equilibria-condition}.
In practice, we define a convergence criterion as a function
of the current and next iterate's states:
\begin{equation}
\label{eq:convergence-metric}
\text{Conv}\left(\statebm{1:\horizon}^{\variterations}, \statebm{1:\horizon}^{\variterations-1}\right) = \left\|\statebm{1:\horizon}^{\variterations} - \statebm{1:\horizon}^{\variterations-1}\right\|_\infty.
\end{equation}
We compute $\statebm{1:\horizon}^{\variterations}$ based on the proposed controls resulting from update step \cref{eq:silq-update-eq}. 
We say \ac{SILQGames} converges if the metric value falls below a threshold $\tau$. \ac{SILQGames} stops after a maximum number of iterations $M_{\text{iter}}$, irrespective of convergence. We expect \ac{SILQGames} to converge, though we do not expect monotone decrease in the convergence criterion as a large step size may occasionally overshoot the Stackelberg equilibrium. 
Oscillations in the convergence metric can occur when step sizes are consistently too large and may indicate that $\alpha_{\min}$ or $\beta$ should be reduced.
Please refer to our results in \cref{ssec:silqgames-experiments-results} for further details.

\noindent\textbf{Computational Complexity.} The complexity analysis of \cite{fridovich2020efficientiterative} holds almost identically for \ac{SILQGames}. \edit{
For a size-$\numstates$ state, linearizing the dynamics and computing quadratic cost approximations both require taking $O(\numstates^2)$ partial derivatives. Solving the coupled Ricatti equations for the approximate LQ game has complexity $O(n^3)$ for a constant ($\numplayers = 2$) number of agents, so }the per-iteration runtime of \ac{SILQGames} is \edit{cubic in $\numstates$.} 
%
The entire algorithm runs in $O(\variterations\numstates^3)$, where $\variterations \! \leq \! M_{\text{iter}}$ is the number of iterations to convergence.

\newcommand\allctrls[1]{w_{#1}}
\newcommand\allctrlsbm[1]{\bm{w}_{#1}}

\subsection{Leadership Filtering}
\label{ssec:leadership-filtering}
The \acl{SLF} (\ac{SLF}) estimates the likelihood that each agent is the leader of a two-agent interaction given noisy measurements $\statemeasbm{1:\horizon}$. Let filter context $\context{t} = [\state{t}, \hyp{t}]^\T$ consist of continuous game state $\state{t}$ and leader $\hyp{t}$.
Following conventional Bayesian filtering practices and denoting all agent controls $\allctrls{t} \!=\! \{ \ctrl{(1)}{t}\!\!, \ctrl{(2)}{t} \}$ for brevity, the \ac{SLF} refines prior context belief $\beliefc{t-1}$ with update rule
\begin{flalign}
\label{eq:belief-update}
\beliefc{t} \propto \prob{\statemeas{t} | \state{t}} \int_{\context{t-1}} \!\!\!\!\! \probl{\context{t} | \context{t-1}, \allctrls{t-1}} \beliefc{t-1} d\context{t-1},
\end{flalign}
$$\vspace{-18pt}$$
In \cref{eq:belief-update}, the context transition probability term $\prob{\context{t} | \context{t-1}, \allctrls{t-1}} = \prob{\state{t}, \hyp{t} | \state{t-1}, \hyp{t-1}, \allctrls{t-1}}$ describes the likelihood of context $\context{t}$ given the previous context $\context{t-1}$ and each agent's controls. Furthermore, the measurement likelihood $\prob{\statemeas{t} | \state{t}}$ quantifies an expected measurement based on how well the new state $\state{t}$ matches the observation $\statemeas{t}$. Thus, we compute the leadership belief at time $t$ by marginalizing $\beliefc{t} = b(\state{t}, \hyp{t})$ over $\state{t}$:
\vspace{-4pt}
\begin{flalign}
\label{eq:leadership-extraction}
\belief{t} = \int_{\state{t}} \beliefboth{t} d\state{t}.
\end{flalign}
$$\vspace{-20pt}$$

To simplify the context transition probability, we assume conditional independence of $\state{t}$ and $\hyp{t}$ given $\context{t-1}$ and $\allctrls{t-1}$. While these values often evolve together, we can make this assumption if the state responds slowly to changes in leadership. In particular, if we select a sufficiently small sampling period $\Delta t$, then changes in state $\state{t}$ when $\hyp{t} \neq \hyp{t-1}$ require multiple time steps to observe. 
After this simplification, 
\begin{equation}
\label{eq:lf-assumption-1}
\!\!\!\!\!\resizebox{0.92\columnwidth}{7.5pt}{
$\probl{\context{t} | \context{t-1}, \allctrls{t-1}} \!=\! \probl{\state{t} | \context{t-1}, \allctrls{t-1}} \probl{\hyp{t} | \context{t-1}, \allctrls{t-1}}$.}
\end{equation}
The term $\prob{\state{t} | \context{t-1}, \allctrls{t-1}}$ indicates that $\state{t}$ depends on the previous leader and the previous state and controls through the dynamics $f_{t-1}$. The second term $\prob{\hyp{t} | \context{t-1}, \allctrls{t-1}}$ \edit{describes} how $\hyp{t}$ depends on the previous state and controls. \edit{In the passing scenario, for example, analyzing this term might allow us to test for a relationship between $\hyp{t}$ and whichever vehicle was in front at time $t-1$.}

\edit{Constructing the \ac{SLF} as a Bayesian filter first requires a leadership transition process. However, establishing a form} for the $\prob{\hyp{t} | \context{t-1}, \allctrls{t-1}}$ term is difficult \cite{strandburg2018inferring,garland2018anatomy}, so we leave it to user discretion if such knowledge is available. \edit{In the case that a form does not exist, we 
treat the leadership transition process as a two-state Markov Chain with transition likelihood $\prob{\hyp{t} \neq \hyp{t-1} | \hyp{t-1}} = p_{\text{trans}}$. In this chain, $\hyp{t}$ evolves independently of state $\state{t-1}$ and agent controls $\allctrls{t-1}$. One example in which this treatment appropriately models leadership is when leadership is correlated with distraction and can thus be modeled as only dependent on time. Despite this simplification in construction, our experiments show the \ac{SLF} still 
accounts for the statistical dependence between leadership, state, and controls.}
\begin{figure*}
\centering
\centering
\begin{minipage}{0.45\textwidth}
    \centering
    \begin{subfigure}[t]{\columnwidth}
        \centering
        \vspace{10pt}
        \includegraphics[width=\columnwidth]{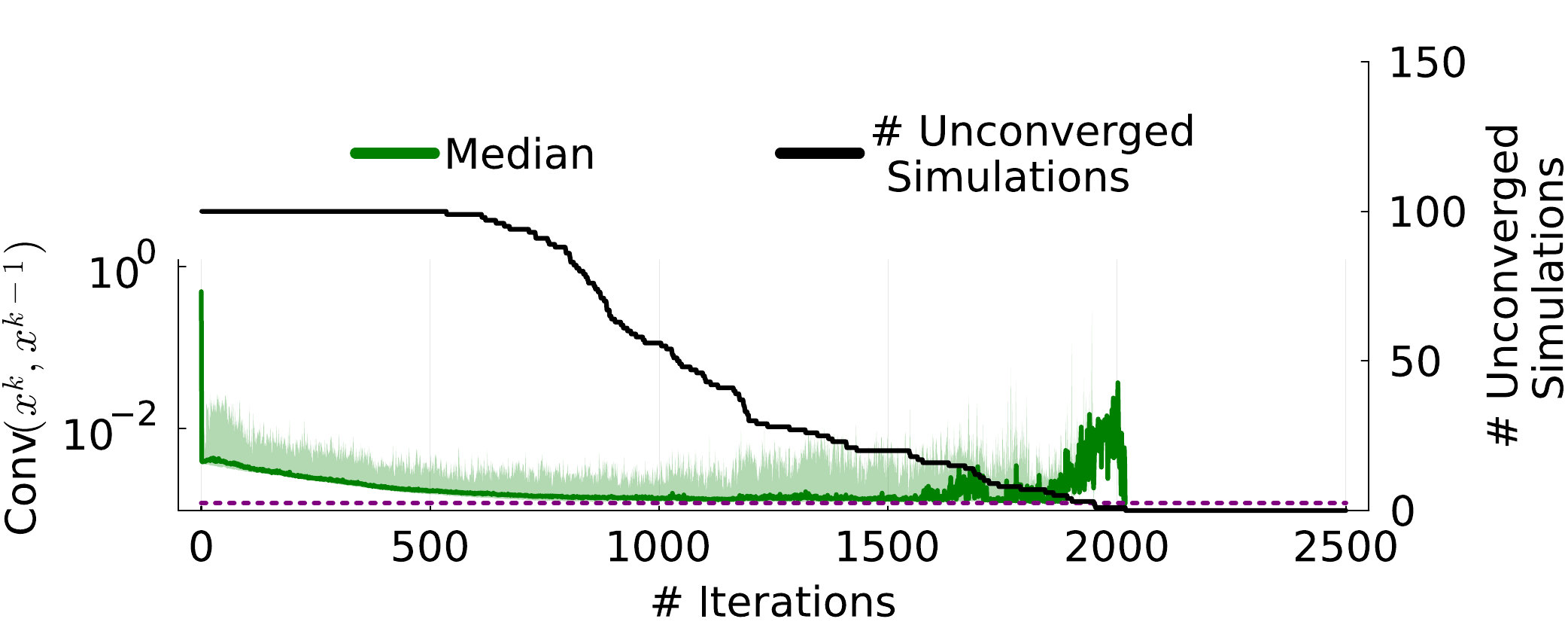}
        \caption{Median $\ell_\infty$ convergence metric, with 10th and 90th percentiles.
        }
        \label{fig:mc-nonlq-silqgames-convergence}
    \end{subfigure}
\end{minipage}
\hfill
\begin{minipage}{0.25\textwidth}
    \vspace{6pt}
    \begin{subfigure}[t]{\columnwidth}
        \centering
        \includegraphics[width=\columnwidth]{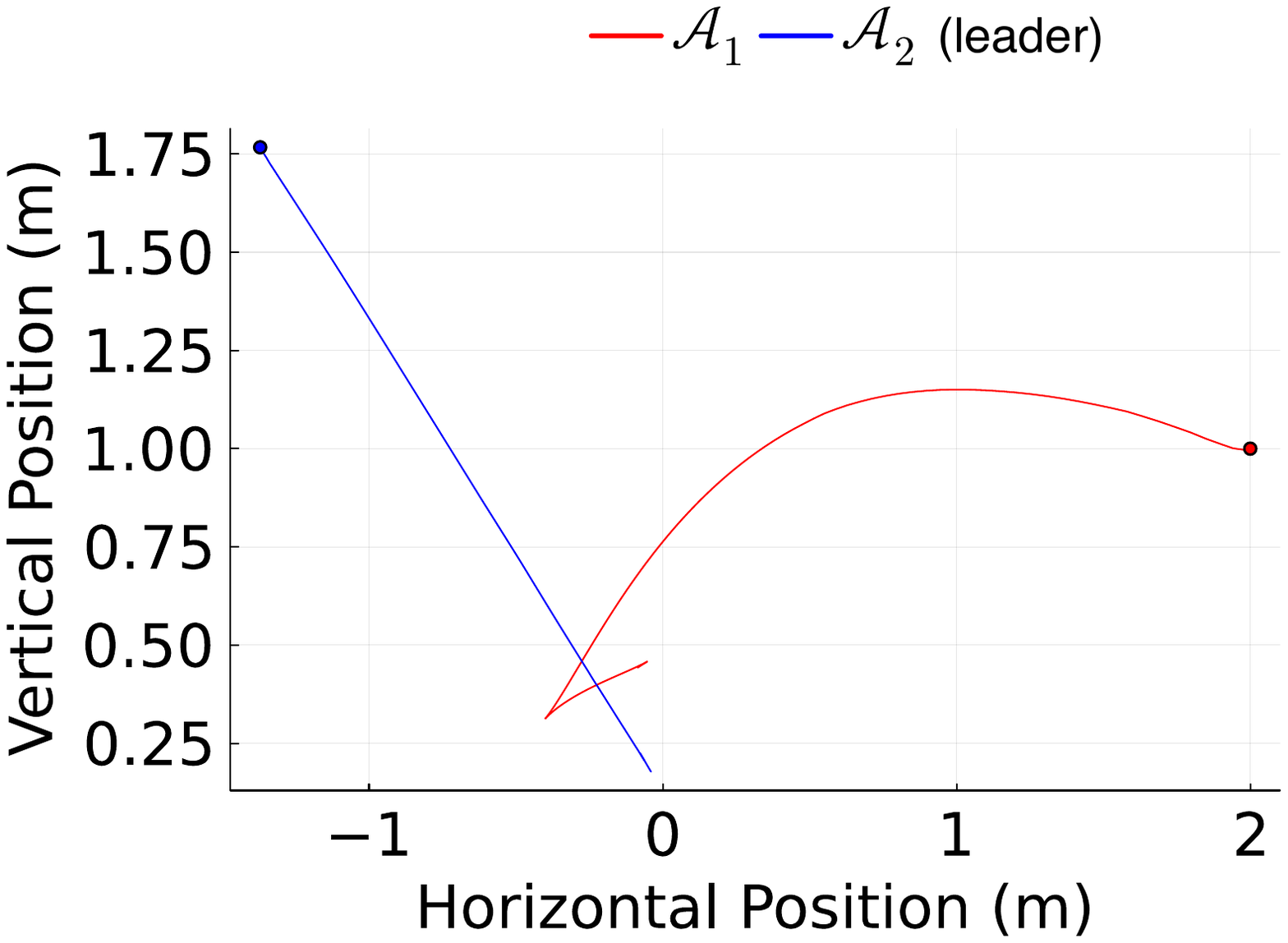}
        \caption{Stackelberg solution positions.}
        \label{fig:silqgames-nonlq-leader-2-position}
    \end{subfigure}
\end{minipage}
\hfill
\begin{minipage}{0.2\textwidth}
    \caption{We run \NonLQNumSims{} \ac{SILQGames} simulations on the non-\ac{LQ} shepherd and sheep game \edit{with leader $\agent{2}$}. 
    The simulations converge in \edit{$1133 \pm 367$} 
    iterations. (a) shows the number of unconverged simulations and (b) shows the solution for one instance.
    }
    \label{mc-nonlq-silqgames-results}
\end{minipage}
\vspace{-15pt}
\end{figure*}

\noindent\textbf{Selecting a Filter.} 
Due to the computational intractability of exactly evaluating Bayesian update rule \cref{eq:belief-update}, we use a particle filtering approach.
Particle $k$ has context $\contextk{t}{k} = [\particlek{t}{k}, \hypk{t}{k}]^\T$.
Particle filters use a measurement model to compute the expected observation for a state $\state{t}$ \cite[Ch. 4]{thrun2002probabilistic}. Our measurement model $h(\particlek{t}{k}; \hypk{t}{k})$ solves a Stackelberg game to generate simulated solution trajectories conditioned on the particle's leader. In the measurement update, we compare a subset of the solution to the ground truth observations 
and update the likelihood of leadership using \cref{eq:belief-update,eq:leadership-extraction}. We resample with replacement to eliminate unlikely particles when the effective number of particles, a metric that measures how well the particles represent the distribution, becomes low. We infer the leading agent based on the similarity of expected measurements, generated from Stackelberg games, to observations of the ground truth.
%
%
Since Stackelberg equilibria satisfy leadership condition \cref{eq:stack-leader-equilibria-condition}, converged solutions let the filter observe leadership indirectly via the measurement model. 

\noindent\textbf{The Stackelberg Measurement Model.}
\label{sssec:stackelberg-meas-model}
We construct a measurement model that relates the leader $H_{t-1}^k$ in particle $k$ at time $t\!-\!1$ with the expected state measurement at time $t$; in particular, we model the expected measurement from each particle as \edit{an equilibrium strategy of}
game $\stack{\hypk{t-1}{k}}{T_s}(\particlek{t-1}{k})$.
%
\edit{We solve this game with \ac{SILQGames} over horizon $T_s$, taking the initial state and leader from previous particle context $\contextk{t-1}{k}$.}

\edit{For the third input, we require the user to provide an application-specific function to specify nominal strategies $\ctrlbm{(i), \variterations{}}{t-1:t+T_s-1}$ using previous particles, a heuristic, etc. The \ac{SLF} calls this function within the measurement model to produce nominal strategies as input to \ac{SILQGames}.
We describe one such heuristic in the appendix.
We call the solutions to these games \emph{Stackelberg measurement trajectories} and select the state at time $t$
as the expected measurement.}
%
%

Next, we clarify a few practical details.
First, experiments determine that we must configure $T_s$ carefully, neither too short to provide relevant leadership information nor too long as to cause excessive latency. 
Second, playing a Stackelberg game from previous state $\state{t-1}$ requires each particle to maintain $\state{t-1}$ as additional context.
Third, after producing a measurement trajectory, we attach measurement uncertainty $\cov{t}$ to each state in it. Depending on the application, this step may incorporate uncertainty from sensors, processing, etc.

\section{Experiments \& Results}
\label{sec:experiments-and-results}

We first introduce the two-agent \ac{LQ} shepherd and sheep game \cite{laine2021shepherdsheepgame} and a nonlinear, nonquadratic variant, \edit{which} we use to validate \ac{SILQGames} and the \ac{SLF}. Finally, we run the \ac{SLF} on realistic driving scenarios.

In the \edit{\ac{LQ}} shepherd and sheep game, \edit{each agent's state $\state{t}^{(i)}$ evolves according to planar double-integrator dynamics \cite[Eq. 75]{taylor2022dynamicsmodels} discretized at $\Delta t$.}
The game state combines the agent states $\state{t} = [\state{t}^{(1)}, \state{t}^{(2)}]^\T$, \edit{and each agent controls its horizontal and vertical accelerations}. Agents' costs
\begin{flalign}
&\hspace{-7pt}g_t^{(1)}\left(\state{t}, \ctrl{(1)}{t}, \ctrl{(2)}{t}\right) = \left(p^{(2)}_{x, t}\right)^2 + \left(p^{(2)}_{y, t}\right)^2 + \|\ctrl{(1)}{t}\|_2^2, \label{eq:shepherd-sheep-costs-p1}\\
&\hspace{-7pt}g_t^{(2)}(\ldots) = \left(p^{(1)}_{x, t}-p^{(2)}_{x, t}\right)^2 \hspace{-2pt} + \left(p^{(1)}_{y, t} - p^{(2)}_{y, t}\right)^2 \hspace{-2pt} + \|\ctrl{(2)}{t}\|_2^2,
\label{eq:shepherd-sheep-costs-p2}
\end{flalign}
are 
quadratic in state and controls and
incentivize ``shepherd'' $\agent{1}$ to minimize ``sheep'' $\agent{2}$'s distance to the origin (i.e., the barn) and $\agent{2}$ to minimize its distance to $\agent{1}$. \edit{We denote the planar positions for $\agent{i}$ as $p^{(i)}_{x,t}, p^{(i)}_{y,t}$.}
An analytic Stackelberg solution exists since the game is \ac{LQ}. \edit{When framing the shepherd and sheep game as a Stackelberg game, we note that either agent can be selected as the leader.}

\newcommand{\yawrate}[2]{\omega^{#1}_{#2}}
\newcommand{\accel}[2]{\alpha^{#1}_{#2}}

We form a nonlinear, nonquadratic variant of \cref{eq:shepherd-sheep-costs-p1}, \cref{eq:shepherd-sheep-costs-p2} by using planar unicycle dynamics \edit{\cite[Eq. 77] {taylor2022dynamicsmodels} with a velocity state that evolves according to $\dot{v} = \alpha$. We discretize the dynamics at $\Delta t$.} Each agent $\agent{i}$ controls yaw rate $\yawrate{(i)}{t} \in \mathbb{R}$ and longitudinal acceleration $\accel{(i)}{t} \! \in \! \mathbb{R}$.
%
The nonquadratic cost
\begin{flalign}
\label{eq:shepherd-sheep-lnq-costs}
g_t^{(1')}&\left(\state{t}, \ctrl{(1)}{t}, \ctrl{(2)}{t}\right) = g_t^{(1)}(\cdot,\cdot,\cdot) - \log\left(\ell - p^{(2)}_{x, t}\right) \\ &\hspace{5pt}- \log\left(p^{(2)}_{x, t}  - \ell\right) - \log\left(\ell - p^{(2)}_{y, t}\right) - \log\left(p^{(2)}_{y, t} - \ell\right) \nonumber
\end{flalign}
adds log barrier terms to \cref{eq:shepherd-sheep-costs-p1} which force $\agent{1}$ to keep $\agent{2}$'s position $(p^{(2)}_{x,t}, p^{(2)}_{y,t})$ bounded within an origin-centered square of side length $2\ell$. The cost remains convex.


%
%

\subsection{SILQGames Validation}
\label{ssec:silqgames-experiments-results}

To test convergence for non-\ac{LQ} games, we run \NonLQNumSims{} simulations of \ac{SILQGames} on the non-\ac{LQ} shepherd and sheep game \edit{with $\agent{2}$ as leader}. We fix $\agent{1}$'s initial position at $(\SI{2}{\meter}, \SI{1}{\meter})$ and vary $\agent{2}$'s initial position along the perimeter of a radius-$\SI{\sqrt{5}}{\meter}$ circle. Both agents begin stationary and face toward the origin. The nominal strategies apply zero input. 
We specify additional parameters in the appendix.

\begin{figure*}
\centering
\begin{minipage}{0.33\textwidth}
    \centering
    \vspace{10pt}
    \begin{subfigure}[t]{\textwidth}
        \includegraphics[width=0.9\columnwidth]{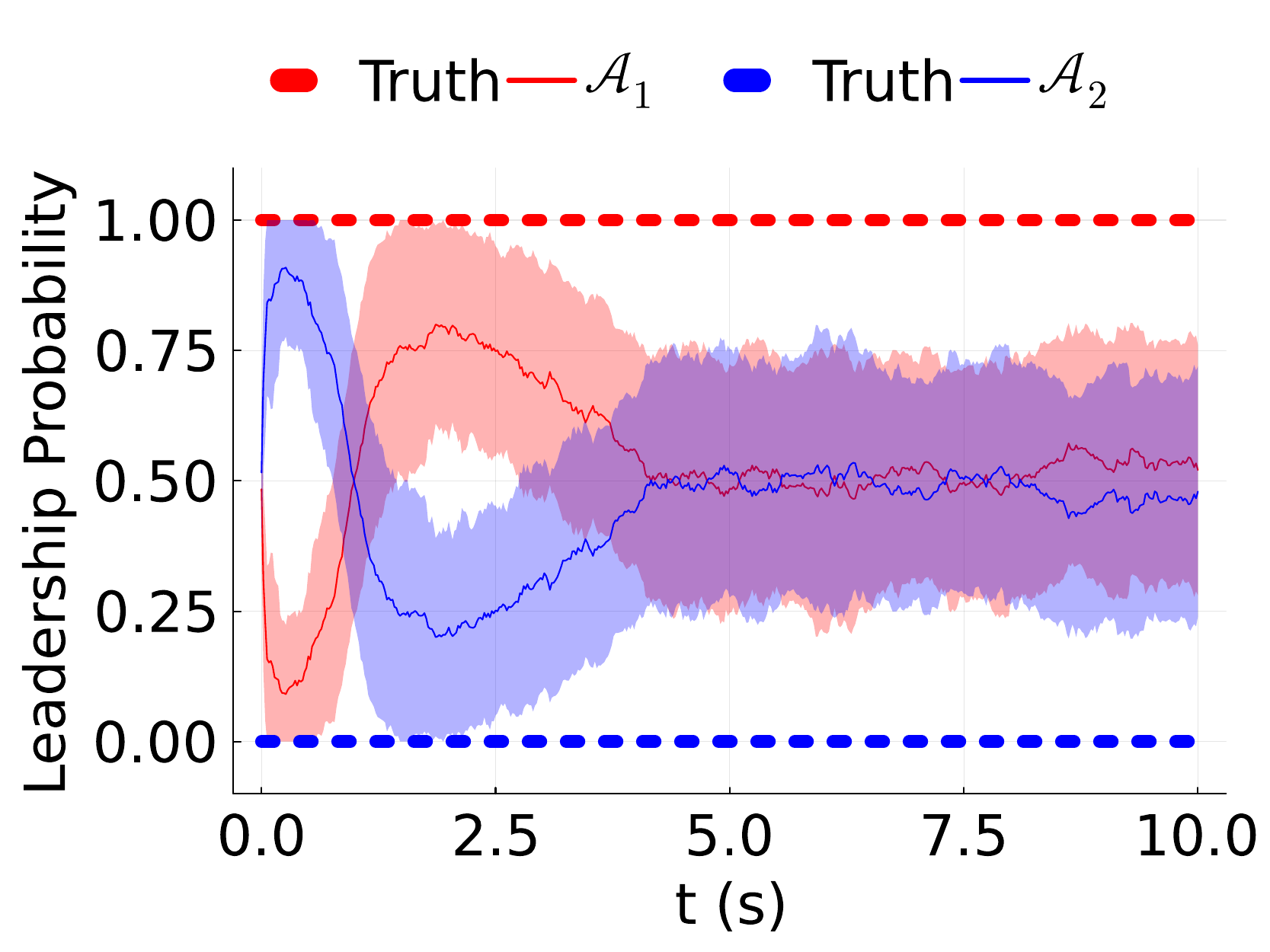}
        \subcaption{\vspace{0pt}Mean leader probability with standard deviation.}
        \label{fig:mc-lf-lq-leader-1-prob-1}
    \end{subfigure}
\end{minipage}
\hfill
\begin{minipage}{0.65\textwidth}
    \centering
    \vspace{2pt}
    \hspace{0.175\columnwidth}\includegraphics[width=0.75\columnwidth]{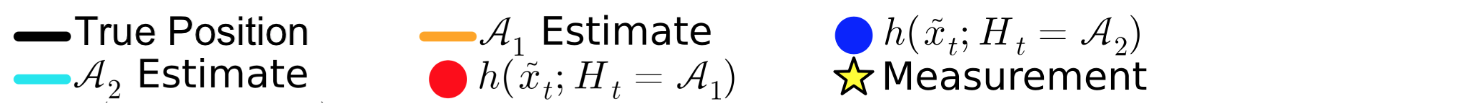}
    \vspace{-12pt}
    \begin{multicols}{2}
        \begin{subfigure}[t]{\columnwidth}
            \centering
            \includegraphics[width=\columnwidth]{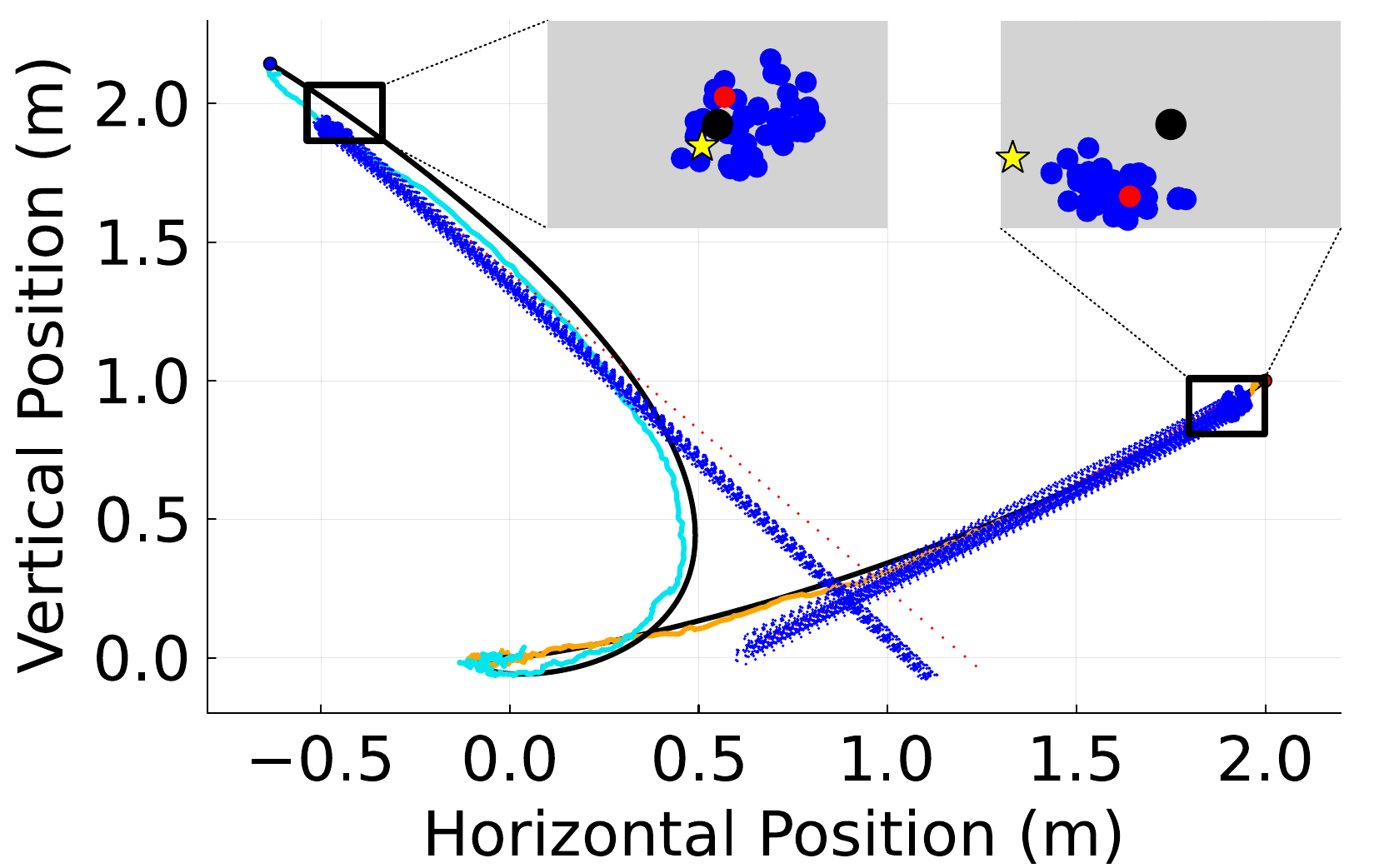}
            \subcaption{Measurement trajectories at $t=\SI{0.54}{\second}$.}
            \label{fig:leadership-filter-lq-leader-1-detail-t27}
        \end{subfigure}
        \\
        \begin{subfigure}[t]{\columnwidth}
            \centering
            \includegraphics[width=\columnwidth]{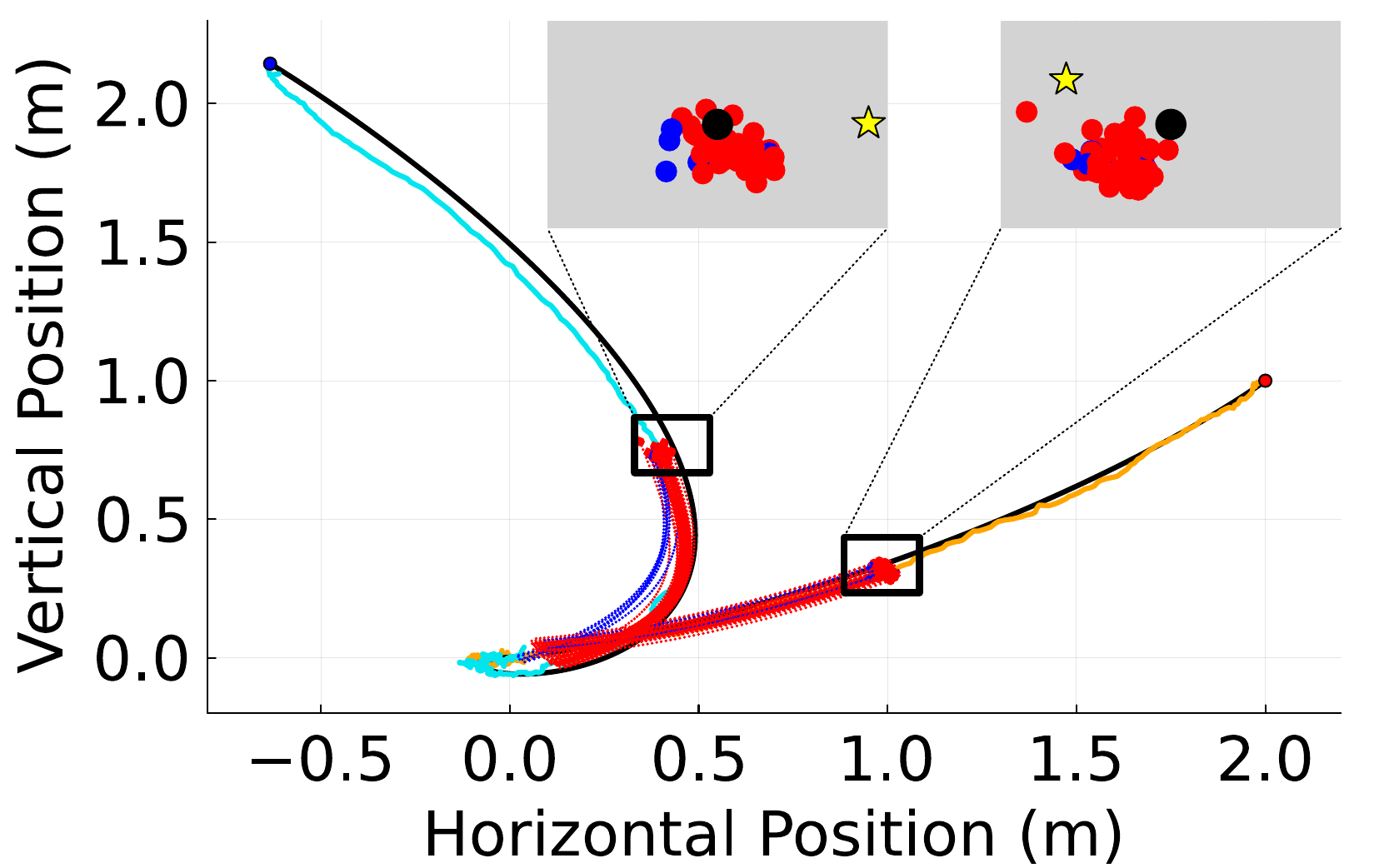}
            \subcaption{Measurement trajectories at $t=\SI{2.04}{\second}$.
            }
            \label{fig:leadership-filter-lq-leader-1-detail-t102}
        \end{subfigure}
    \end{multicols}
\end{minipage}
\caption{We run \NonLQNumSims{} \ac{SLF} simulations on analytic solutions to the \ac{LQ} shepherd and sheep game.
(a) indicates that the \ac{SLF} initially misidentifies the leader but then identifies the leader correctly as $\agent{1}$ before becoming uncertain due to noise. (b) and (c) are associated with a particular simulation and show the Stackelberg measurement trajectories at $t=\SI{0.54}{\second}$ and $t=\SI{2.04}{\second}$, respectively. The color of a particle's measurement trajectory indicates leading agent $\hypk{t-1}{k}$,
and the insets show the expected measurements for each particle, the actual measurement, and the ground truth.
\vspace{-15pt}
}
\label{fig:leadership-filter-lq-leader-1-results-global}
\end{figure*}

\noindent\textbf{Analysis.} The results in \cref{fig:mc-nonlq-silqgames-convergence} indicate that all simulations converge. The median value of the convergence metric, shown with 10\% and 90\% percentile bounds, exhibits a generally decreasing trend.
These results are consistent with our previous discussion on convergence, as \ac{SILQGames} converges in every simulation, though without monotone decrease in the convergence criterion. \edit{\ac{SILQGames} behaves similarly when we vary the initial position of the follower.}
 
In \cref{fig:silqgames-nonlq-leader-2-position}, we report the solution for a particular (arbitrarily chosen) simulation. Both agents' motion follows the incentive structure of the game: the distance between the two agents decreases, as does the distance from $\agent{2}$ to the origin. As expected, $\agent{1}$ exerts more control effort than $\agent{2}$ due to $\agent{2}$'s leadership role and $\agent{1}$'s incentive to constrain $\agent{2}$'s position.
%
Finally, we note that $\agent{1}$'s motion changes sharply towards the end of its trajectory. Here, the unicycle comes to a stop and moves in reverse. These results demonstrate that, for a game with nonlinear dynamics and convex, nonquadratic costs, \ac{SILQGames} converges to a solution that appears consistent with the dynamics and costs.


\noindent\textbf{Timing.} We collect elapsed times for each iteration of \NonLQNumSims{} \ac{SILQGames} simulations on AMD Ryzen 9 5900x 12-core processors. \edit{The per-iteration runtime (with standard deviation) of \ac{SILQGames} is $0.49 \pm \SI{0.29}{\second} $.} \edit{We note that straightforward but nontrivial optimizations (i.e., more principled step size selection, numerical optimization techniques, etc.) such as those used in other iterative game solvers \cite{peters2020inference,lasse2020mastersthesis} have been shown to improve computational efficiency.
}



\subsection{Leadership Filter Validation}
\label{ssec:leadership-filter-experiments-results}

We validate the leadership filter on analytic solution trajectories of horizon $T_{\text{sim}}$ for the \ac{LQ} shepherd and sheep game played with leader $L_{\GT} \!=\! \agent{1}$.
Since we generate the ground truth $\statebm{1:\horizon}^{\GT}$ with a known leader, a perfect filter should infer the true leader with consistently high confidence. 
Our results suggest that the \ac{SLF} produces an observable signal for Stackelberg leadership, but (as one can expect) noise and measurement model configuration significantly affect performance.
We simulate noisy state measurements $\eststate{t} \!\sim\! \mathcal{N}(\state{t}^{\GT}\!, \cov{})$
and pass $\cov{}$ to the \ac{SLF}.
We list parameter values in the appendix.



\newcommand{\Tone}{27}
\newcommand{\Ttwo}{102}
\newcommand{\Tthree}{352}

\newcommand{\LQInstanceNum}{100}

\noindent\textbf{Analysis.} 
In our results, the \ac{SLF} produces the expected leadership probability
for part of the simulation horizon. 
%
%
From $1.5-\SI{3.5}{\second}$ in \cref{fig:mc-lf-lq-leader-1-prob-1}, the \ac{SLF} correctly infers $\agent{1}$ as the leader with high likelihood.
%
Examining the expected measurements in \cref{fig:leadership-filter-lq-leader-1-detail-t102} at $\SI{2.04}{\second}$, 
we note that 
the observations in this time range more closely match the measurement models generated with $\agent{1}$ as leader, which the \ac{SLF} interprets as indicating leadership by $\agent{1}$. 
\edit{In the left inset of \cref{fig:leadership-filter-lq-leader-1-detail-t102}, the particles that believe $\agent{2}$ to be the leader have positions further to the left of those that consider $\agent{1}$ to be the leader. From this effect, we see that the \ac{SLF} accounts for the statistical dependence of state and leadership despite our simplified treatment of the leadership transition process.}

%

However, we also see complex behavior in \cref{fig:mc-lf-lq-leader-1-prob-1}. First, the \ac{SLF} initially misidentifies the leader as $\agent{2}$, as shown by \cref{fig:leadership-filter-lq-leader-1-detail-t27}, because the Stackelberg measurement trajectories do not capture leadership information over the whole simulation horizon.
%
Specifically, the measurement trajectories $\{ h(\particlek{t-1}{k}, \hypk{t-1}{k}) \}$ are straight lines that roughly reduce the state costs of the shepherd and sheep, but do not capture the granularity of motion from the ground truth due to higher control costs over the short horizon $T_s \ll T_{\text{sim}}$.

Second, the \ac{SLF} is completely uncertain after $\SI{4.5}{\second}$. 
Near the origin, the contribution of process noise to the motion outweighs the contribution of the dynamics, and together with measurement noise
obfuscate the dynamics.
%
%

From these results, we see that the \ac{SLF} requires parameter $T_s$ to be of sufficient length to capture the influence of leadership on the measurement trajectories.
We note that the \ac{SLF} is sensitive to noise as it infers leadership indirectly by comparing the observed motion with the expected motion of a Stackelberg leader. 
Thus, too little process noise may lead particles to converge to an incorrect trajectory, and too much reduces the signal-to-noise ratio.



\noindent\textbf{Timing.} 
The mean overall runtime for \LQNumSims{} simulations of an LQ game with \LQTimeSteps{} steps \edit{is $10.91 \! \pm \! \SI{1.64}{\second}$.}
\edit{For $50$ particles and a $75$-step measurement horizon, each step of the \ac{SLF} runs in $0.82 \pm \SI{0.35}{\second}$.}
Self-driving vehicle applications require sub-\SI{100}{\milli\second} perception cycle computation time \cite{lin2018architectural}, so our implementation is not real-time. \edit{To meet real-time computational efficiency requirements, we must parallelize particle computation and optimize \ac{SILQGames}; the latter is the most expensive step for each particle.}
\edit{These changes} have been shown to reduce computation time below $\SI{100}{\milli\second}$, as demonstrated by \cite{peters2020inference,lasse2020mastersthesis}, which use fast particle filters with measurement models that solve dynamic games.
\newcommand{\initstate}[1]{\bm{x}^{#1}_0}
\newcommand{\vmax}{35}
\newcommand{\passingTs}{1.0}
\newcommand{\vinit}{10}
\newcommand{\dxinit}{10}
\newcommand{\dc}{0.2}
\newcommand{\lb}{2.5}
\newcommand{\maxaccel}{9}
\newcommand{\maxrotvel}{2}
\newcommand{\leadprior}{0.5}
\newcommand{\cl}{0}
\subsection{Realistic Driving Scenarios}
\label{ssec:driving-scenarios}
We formulate passing and merging scenarios using realistic ground truth trajectories without a clear leader. 
We demonstrate that the \ac{SLF} responds to changes in leadership, handles objectives that imperfectly model agent behavior,
and that the results match right-of-way expectations.
The dynamics and cost terms demonstrate that the \ac{SLF} does not require \ac{LQ} assumptions and works for nonconvex costs. Our results further indicate that \ac{SILQGames}, used within the \ac{SLF}, converges under these conditions.


Each agent's state evolves according to unicycle dynamics. The simulation runs for $\horizon$ steps at period $\Delta t = \SI{0.05}{\second}$.
%
We model stage cost $g^{(i)}_t$ as a weighted sum of incentives $g^{(i)}_{j, t}$,

\vspace{-6pt}
\begin{equation}
\label{eq:running-cost-at-time-t}
g^{(i)}_{t} = \sum^{M^{(i)}}_{j=1} w^{(i)}_j g^{(i)}_{j, t}.
\vspace{-6pt}
\end{equation}
Weights $\{w^{(i)}_j\} \subset \reals^+$ specify the relative priorities of subobjectives.
We define $M^{(i)} = 6$ terms to incentivize driving behaviors corresponding to legal or safety considerations.
\begin{subequations}
\label{eq:example-constraints}
\begin{flalign}
    & \hspace{-7pt}g^{(i)}_{1, t} = \text{d}(\state{t}^{(i)}, \state{t}^{(i),\text{goal}}) \label{eq:subobj-1} \\
    & \hspace{-7pt}g^{(i)}_{2, t} = -\log(\| p^{(i)}_t - p^{(j)}_t \|^2_2 - d_c) ~~ \forall i \neq j \label{eq:subobj-2} \\
    & \hspace{-7pt}g^{(i)}_{3, t} = -\log(v_{m} - |v^{(i)}_t|) - \log(\Delta\psi_m - |\psi^{(i)}_t - \psi_\text{r}|) \label{eq:subobj-3} \\
    & \hspace{-7pt}g^{(i)}_{4, t} = (\omega^{(i)}_t)^2 + (\alpha^{(i)}_t)^2 \label{eq:subobj-45} \\
    & \hspace{-7pt}g^{(i)}_{5, t} = -\log(\| p^{(i)}_{t, \text{llb}} - p^{(i)}_t \|^2_2) -\log(\| p^{(i)}_{t, \text{rlb}} - p^{(i)}_t \|^2_2) \label{eq:subobj-6} \\
    & \hspace{-7pt}g^{(i)}_{6, t} = \text{exp}(-(1/2) (p^{(i)}_{t, \text{cl}} - p^{(i)}_t)^\T C^{-1} (p^{(i)}_{t, \text{cl}} - p^{(i)}_t)) \label{eq:subobj-7}
\end{flalign}
\end{subequations}

\noindent Eq. \cref{eq:subobj-1} requires a small distance between vehicle state $\state{t}
^{(i)}$ and goal state $\state{t}^{(i),\text{goal}}$. For this scenario, d$(\cdot, \cdot)$ is a weighted Euclidean distance. Eq. \cref{eq:subobj-2} requires a minimum safety radius $d_c$ between the vehicles. Eq. \cref{eq:subobj-3} requires obeying speed limit $v_{m}$ and avoiding excessive heading deviation $\Delta\psi_m$ from road direction $\psi_{\text{r}}$. Eq. \cref{eq:subobj-45} incentivizes low control effort. Eq. \cref{eq:subobj-6} enforces left and right lane boundaries $p^{(i)}_{t, \text{llb}}, p^{(i)}_{t, \text{rlb}}$, based on lane width $\ell_w$. Eq. \cref{eq:subobj-7} uses a (nonconvex) Gaussian function with covariance $C$ to discourage crossing the center line $p^{(i)}_{t, \text{cl}}$. 
We specify these parameter values in the appendix. Lastly, we define the direction of motion as the $y$-direction and the transverse direction as the $-x$-direction to maintain a righthand coordinate frame.



\newcommand{\TsimPassing}{7.5}
\newcommand{\headingdeviation}{\pi/3}

\noindent\textbf{Passing Scenario.} The passing scenario begins with $\agent{2}$
behind $\agent{1}$ and runs for $\SI{\TsimPassing{}}{\second}$.
In the ground truth trajectories (\cref{fig:passing-diagram}), $\agent{2}$ initially follows $\agent{1}$ for $\SI{2.5}{\second}$, then passes in the other lane, and ends ahead of $\agent{1}$ in the initial lane.
$\agent{1}$ drives along the lane at a constant velocity, applying no controls.

We simulate the leadership filter on the passing maneuver.
%
We expect $\agent{1}$ to start with a high leadership probability and for that probability to decrease once the passing maneuver begins, and vice versa for $\agent{2}$. 
%
In \cref{fig:passing-diagram}, the state estimate tracks the ground truth, indicating that the leadership filter captures the game dynamics.
Since the \ac{SLF} produces the expected trends in the state estimates and agents' probabilities,
%
%
our results show that Stackelberg leadership can match right-of-way expectations for scenarios without a ground truth leader. Moreover, the \ac{SLF} responds appropriately to changing leadership dynamics over time. 

Lastly, we analyze the computation time of the \ac{SLF} ($0.027 \pm \SI{0.03}{\second}$ per particle, per step). We note that calls to \ac{SILQGames} converge in $4.2 \pm 13.6$ iterations. During the straight portions of the passing maneuver, the nominal trajectories produce better \ac{LQ} approximations and thus \ac{SILQGames} converges faster. During the turns, poor nominal strategies lead to slower convergence and result in variability in the computation time of the \ac{SLF}.
Overall, our results show that \ac{SILQGames} can handle nonconvex cost terms.


\newcommand{\lengthTwoLanes}{30}
\newcommand{\lengthMerge}{30}
\newcommand{\laneWidthMerge}{\lb{}}
\newcommand{\widthBeginMerge}{5}

\newcommand{\TsimMerging}{5.0}
\newcommand{\mergingTs}{1.0}

\begin{figure}[!ht]
\centering
\vspace{6pt}
\includegraphics[width=0.85\columnwidth]{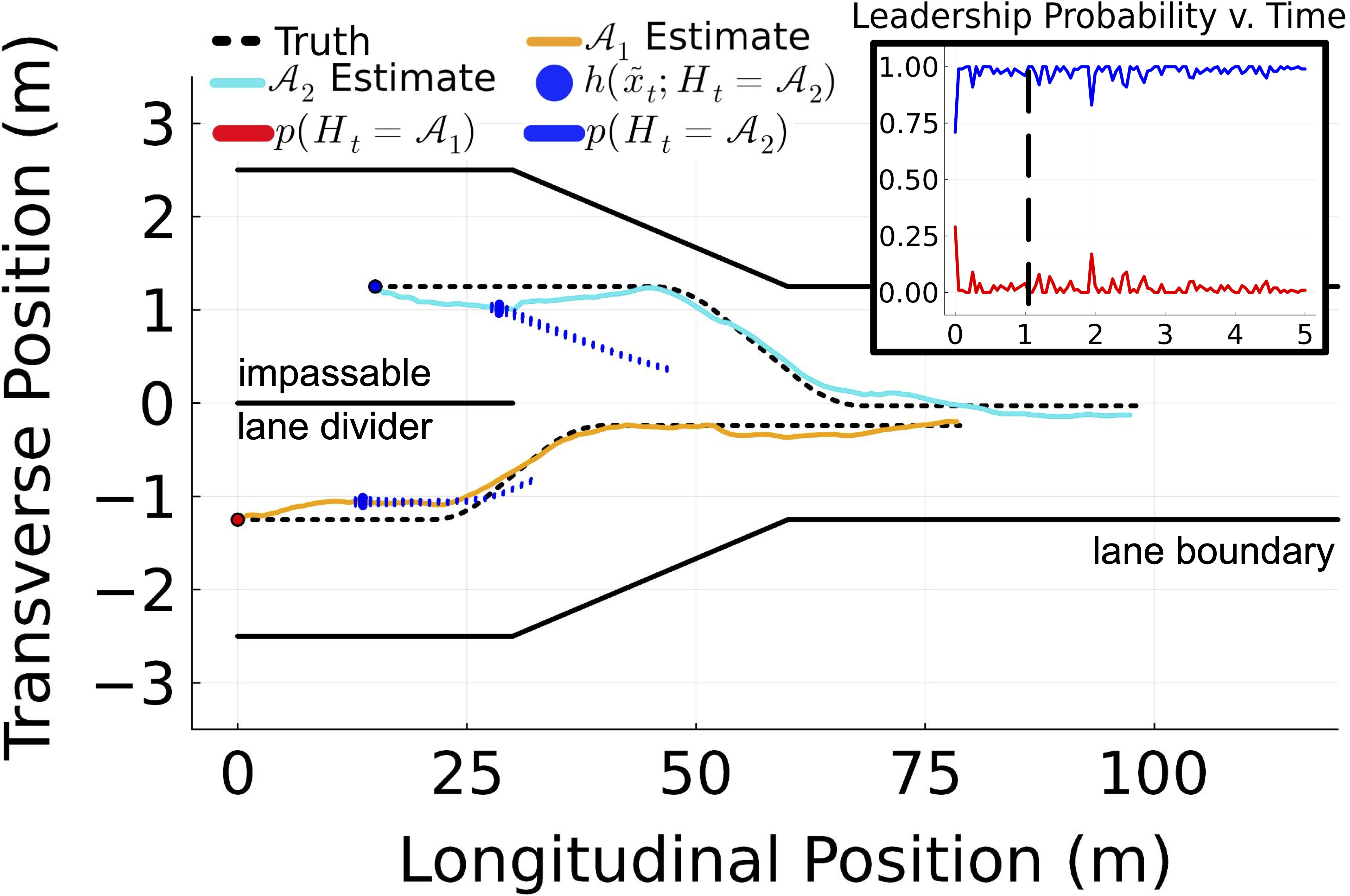}
\caption{In this merge, $\agent{2}$ starts ahead in its lane and $\agent{1}$ yields to $\agent{2}$. We see a high leadership likelihood for $\agent{2}$, as expected because it merges first.
The inset indicates the current probabilities with a vertical dashed line.
\vspace{-12pt}
}
\label{fig:merging-scenario-2-results}
\end{figure}

\noindent\textbf{Merging Scenario.}
The merging scenario involves three sections of road (see \cref{fig:merging-scenario-2-results}): two $\SI{\lengthTwoLanes{}}{\meter}$-long lanes separated by a barrier at $x=\SI{0}{\meter}$, a merging segment that decreases from width $2 \ell_w$
to $\ell_w$
over $\SI{\lengthMerge{}}{\meter}$ of length, and a one lane road centered along $x=\SI{0}{m}$.
Both agents start in their own lanes, though $\agent{1}$ starts behind $\agent{2}$. In the ground truth, $\agent{2}$ merges before $\agent{1}$, which slows down to yield before merging. $\agent{2}$ delays its merge once it enters the merging segment. We construct the game played within the measurement model to incentivize each agent to merge quickly after entering the merging segment, so the cost we define for $\agent{2}$ does not exactly reflect its actual behavior.



In \cref{fig:merging-scenario-2-results}, we simulate this merge with the leadership filter. We expect $\agent{2}$ to lead the interaction as it begins ahead and merges first.
%
Given their objectives, we expect the agents' measurement trajectories to merge quickly, and we see these trajectories quickly move toward the center of the merging segment. 
Nevertheless, the leadership filter's state estimate tracks the ground truth, including $\agent{2}$'s delayed merge, and 
the \ac{SLF} infers $\agent{2}$ as the leader.
%
%
%
Thus, the results match our right-of-way expectations despite agent objectives that do not exactly describe the observed ground truth behavior. \edit{This mismatch results in poor initialization, which affects the computation time ($1.02 \pm \SI{0.99}{\second}$ per particle, per step) as calls to \ac{SILQGames} take longer to converge ($69.2 \pm 38.3$ iterations). As with the passing scenario, the variance in computation time reflects a degradation in the quality of the \ac{LQ} approximation. 
We further re-iterate that parallelization is critical for the \ac{SLF} to run in real-time.}



\section{Discussion \& Limitations}


We contribute \ac{SILQGames}, an iterative algorithm to solve Stackelberg games with nonlinear dynamics and nonquadratic costs. Through empirical validation on non-\ac{LQ} game scenarios, we show it reliably converges.
We also introduce the \acl{SLF} and apply it to noisy scenarios with known leaders and realistic driving situations. Results highlight the \ac{SLF}'s ability to estimate leadership in long-horizon interactions with changing leadership and with objectives that do not exactly reflect observed agent behavior. Furthermore, we discuss 
the robustness of our method 
to the measurement horizon and noise.

Future directions include extending \ac{SILQGames} to $\numplayers > 2$ agents and overcoming combinatorial scaling challenges \edit{arising} from the pairwise definition of Stackelberg leadership. \edit{The number of possible $N$-agent Stackelberg hierarchies grows exponentially and resolving the dependencies in any such hierarchy is nontrivial.} Another critical direction involves establishing theoretical bounds on the number of \ac{SILQGames} iterations. For the \ac{SLF}, future work includes enabling real-time application using more efficient estimators and algorithmically adjusting the measurement horizon $T_s$ to observe leadership dynamics over different horizons.



\section*{Acknowledgment} We thank Professor Todd Humphreys and members of the CLeAR and SWARM Labs at UT Austin for feedback.

\printbibliography  

\appendix
\label{appendix}

\noindent \textbf{\ac{SILQGames} Parameters.} We vary the initial position of $\agent{2}$ 
about $(\SI{-1}{\meter}, \SI{2}{\meter})$ 
along a $\SI{\NonLQanglediff{}}{\radian} $ arc of a circle.
We set convergence threshold $\tau = \NonLQConvThresholdSILQ{}$, the maximum number of iterations to $\NonLQMaxItersSILQ{}$, and minimum step size $\alpha_{\min} = \NonLQMinStepSize{}$. We play the game for $\SI{\NonLQHorizon{}}{\second}$ with period $\Delta t = \SI{\NonLQSamplePeriod{}}{\second}$  ($\NonLQTimeSteps{}$ steps). The nominal controls apply zero input.

\noindent\textbf{\ac{SLF} Parameters.} 
%
In our examples, \edit{we select nominal strategies with a simple heuristic that returns $T_s$-length control trajectories for each agent, i.e. at time $t-1$, the nominal strategy for $\agent{i}$ is $[\ctrl{(i)}{t-1} \cdots \ctrl{(i)}{t-1}]$.}
We configure the number of particles $\numparticles = 50$.
%
The Stackelberg measurement horizon $T_s = 75$ steps ($\SI{1.5}{\second}$). Let $p_{\text{trans}} = 0.02$, so transitioning is thus likely enough that particles can switch leadership state and model dynamic leadership transitions without injecting excessive uncertainty into the inference.
For the process noise uncertainty $\procnoise$, we set position and heading variances on the order of magnitude of $10^{-3}$ and velocity variances to $10^{-4}$. \ac{SLF} measurement uncertainty $\cov{} = 5 \cdot 10^{-3} I$. The convergence threshold $\tau=1.5 \cdot 10^{-2}$, the max iteration count $M_{\text{iter}} = 50$, and step size $\alpha_{\min} = 10^{-2}$.


\noindent\textbf{Driving Scenario Parameters.} Let speed limit $v_{m}\! =\! \SI{\vmax{}}{\meter\per\second}$ with initial headings aligned with the road direction $\psi_r$. Lanes are $\ell_w\! =\! \SI{\lb{}}{\meter}$ wide. A safety violation occurs if the vehicles come within $d_c \! =\! \SI{\dc{}}{\meter}$ of one another. We constrain acceleration and rotational velocity magnitudes to $\SI{\maxaccel{}}{\meter\per\second\squared}$ and $\SI{\maxrotvel{}}{\radian\per\second}$.
The measurement horizon $T_s=\SI{\passingTs{}}{\second}$, with sampling periods of $\SI{0.05}{\second}$ ($\SI{20}{\hertz}$). 
We use $100$ particles with equal initial chance of $\agent{1}$ and $\agent{2}$ as leader. 
The center line is at $x = \SI{\cl{}}{\meter}$. Each agent begins with velocity $\SI{\vinit{}}{\meter\per\second}$. Other parameters are identical to the \ac{SLF} parameters.


\end{document}